\documentclass{article}
\usepackage{arxiv}
\usepackage{color}
\usepackage{subfigure}
\usepackage{graphicx}
\usepackage{dcolumn}
\usepackage{bm}
\usepackage{floatrow}
\usepackage{amsmath}
\usepackage{makeidx}
\usepackage{epsfig}
\usepackage{array}
\usepackage{epstopdf}
\usepackage{multirow}
\usepackage[utf8]{inputenc} 
\usepackage[T1]{fontenc}    
\usepackage{hyperref}       
\usepackage{url}            
\usepackage{booktabs}       
\usepackage{amsfonts}       
\usepackage{nicefrac}       
\usepackage{microtype}      
\usepackage{lipsum}	
\usepackage{listofsymbols}
\usepackage{moreverb}
\usepackage{listings} 
\title{A concise introduction to molecular dynamics simulation: Theory and programming}
\author{
  Ashkan Shekaari \\
  Department of Physics\\
  K. N. Toosi University of Technology\\
  Tehran, 15875-4416, Iran \\
  \texttt{shekaari@email.kntu.ac.ir} \\
   \And
 Mahmoud Jafari \thanks{Corresponding author} \\
  Department of Physics\\
  K. N. Toosi University of Technology\\
  Tehran, 15875-4416, Iran \\
  \texttt{jafari@kntu.ac.ir} \\}
\begin{document}
\maketitle
\begin{abstract}
We provided a concise and self-contained introduction to molecular dynamics (MD) simulation, which involves a body of fundamentals needed for all MD users. The associated computer code, simulating a gas of classical particles interacting via the Lennard-Jones pairwise potential, was also written in Python programming language in both top-down and function-based designs.
\end{abstract}
Developed originally by Alder and Wainwright in the 1950s~\cite{1}, and began to gain extensive attention in the mid-1970s concurrent with the advent of powerful computers, molecular dynamics (MD) methods have long been considered as orthodox means for simulating matter at molecular scales, continuing to excite the ardor of researchers with new problems, also from students new to molecular theory as well~\cite{2}. The very essence of MD, simply, is to solve numerically the $N$-body problem of classical mechanics---indeed, solving differential equations is what happens in all physics: Einstein field equations~\cite{3} in general relativity~\cite{4}; Schr{\"o}dinger equation in quantum mechanics~\cite{5}; and Newton's equations of motion in classical mechanics. At present, the importance of such a many-body problem arises from attempts to relate collective behaviors of many-body systems to the associated single-particle dynamics. MD simulation is indeed the modern realization of an old, deterministic, and mechanical interpretation of Nature in the sense that the behavior of the system could be exactly computed if initial conditions, namely positions, velocities, and accelerations of the constituent particles of that system are known~\cite{6}. It has also contributed to much of our understanding of systems out of equilibrium for which, theory, particularly statistical mechanics~\cite{7}, has little to say~\cite{8}.

Not only as a capable scientific method deployed to do research in applied areas of science and technology, such as drug discovery~\cite{9}, biochemistry of diseases~\cite{10}, etc., MD has also been exploited as an educational tool in a number of university courses including computational physics, chemistry, materials science, and biology.

The present work has been devoted to providing a brief, comprehensive, and self-contained overview of MD simulation. Indeed, compared to other topics such as numerical integration or solving differential equations, along with which MD is taught during one semester, the latter requires a dramatically larger amount of educational time due to being inherently more sophisticated and time-consuming. Here, we present a core of fundamentals that should be common to all users of MD. This introduction has been restricted to the simulation of a many-body system, in thermodynamic equilibrium, and composed of Lennard-Jones~\cite{11} atoms as hard spheres. The related computer code was also written in Python language in top-down as well as function-based programming styles. Quantum effects (as in ab initio MD~\cite{12,13,14}), multibody interactions (such as Axilrod-Teller three-body potential~\cite{15}), and simulating nonequilibrium processes~\cite{16} have not been discussed here. Previous exposure to classical and statistical mechanics, as well as programming in Python~\cite{17}, C~\cite{18}, or Fortran~\cite{19}, could also be very useful.
\section{\label{sec:1_1}Primitive concepts}
{\em{System}} is defined as the portion (or subset) of the physical universe on which we concentrate. To investigate the behavior of the system, one needs ways to assign numerical values either to the state or to functions of that state. This assignment is called an {\em{observable}}. As an illustration, the ideal gas law, $PV=Nk_{B}T$, is a relation among the observables pressure $P$, volume $V$, number of particles $N$, and equilibrium temperature $T$ ($k_{B}$ being the Boltzmann constant). The state of the system can be manipulated or changed via interactions with its {\em{environment}}. The system and its immediate environment are separated via system's boundary, which constrains system-environment interactions. Here, by system, we solely mean {\em{isolated system}} the boundary of which does not allow the exchange (entry or exit) of matter and energy with the surroundings.
\subsection{\label{sec:1_3}MD is not a model}
To establish connection between measurable outputs and controlled inputs is the goal of theoretical works. In theory, complicated interactions among state variables are entirely or partially decoupled in order for observable outputs to be computed. A model is indeed a scheme to decouple and eliminate interactions with negligible or no impact on the observables of interest. As a result, a model is simpler than the original system, having then access to fewer states, and vice versa, a model has access to some states which are not available to the system it imitates. A simulation, in contrast, is more complicated than the original system and can accordingly reach more states. However, the original system should not be considered as a model of the simulation at all.
\subsection{\label{sec:1_2}MD simulations are computer experiments}
Controversial arguments have so far arisen over the query of whether computer simulations like MD are theories or experiments. The theory side believe that simulations are not experiments, because they are as such pure calculations and no measurement is done on real systems during a computer simulation. The experiment side, on the other hand, argue that simulations are experiments because their results (i) are used to test theories, (ii) are reproducible, and (iii) are statistically error-prone. Indeed, the latter interpretation is widely accepted and pervaded the literature, and we likewise believe that MD simulations are computer experiments.
\section{\label{sec:2}Fundamentals}
\subsection{Ergodic theorem}
MD is a widespread class of computer simulations used in many areas of science particularly statistical mechanics, and involves two general forms including systems at equilibrium, as well as those away from equilibrium. The reliability of MD as such stems from the ergodic theorem introduced by Boltzmann. One, in fact, needs to be assured that the system to be simulated is ergodic, namely the associated time average and ensemble average are nearly the same, mainly based on the fact that an MD simulation produces a time evolution of that system. Most systems with realistic pairwise potentials fortunately seem to behave ergodically in two or more dimensions---one-dimensional systems should probably be viewed as suspects---in spite of the fact that only a few systems are indeed ergodic. Geometrically, a system is ergodic (or more realistically, quasiergodic) if the trajectory of its representative point (i.e., phase point) crosses any neighborhood of any point of the system's phase space in the course of time. And Mathematically,
\begin{equation*}
\langle A\rangle_{ens}= \frac{\int_{0}^{t_{tot}}dq{\hspace{0.5mm}}A(q){\hspace{0.5mm}}p(q)}{\int_{0}^{t_{tot}}dq{\hspace{0.5mm}}p(q)}\cong\lim_{t_{tot}\longrightarrow\infty}\frac{1}{t_{tot}}\int_{0}^{t_{tot}}dtA(q(t))=\langle A\rangle_{time},
\end{equation*}
meaning that if we wait long enough (i.e., $t_{tot}\longrightarrow\infty$), the ensemble average $\langle A\rangle_{ens}$ and the time average $\langle A\rangle_{time}$ of the thermodynamic observable $A$ of the system would be the same. Here $q$ is a microstate of the system, and might represent the positions and momenta of the constituent particles, and $p(q)$ is the associated probability distribution function. This is indeed based on this theorem that MD simulations are not reliable at very low temperatures due to lack of adequate vibrancy of the phase point---in this so-called low-temperature limit, we have in fact to deploy other methods such a phonon-dispersion calculations to derive the thermodynamics of the system.

In Monte Carlo (MC) methods~\cite{20}, the other vast category of simulations in statistical mechanics, similar averaging is done, but on the phase space of the system, not on the created time-dependent trajectories as in MD. An inevitable error could therefore be assigned to either method: in MC, averaging should be done over the entire phase space, but we are only able to average over a limited number of samples; and in MD, we must wait infinitely, but it is not possible and averaging is practically done over limited durations of time. However, one could minimize such a kind of inherent error via optimally sampling the phase space (in MC), or by accurately estimating the characteristic timestep of the system (in MD).
\subsection{Standard ensemble for MD}
MC is based fundamentally on the Boltzmann weight factor $e^{-\beta\epsilon}$, where $\epsilon$ is the energy of a state of the system, and $\beta=1\big/k_{B}T$ is the inverse temperature of the surroundings. Accordingly, the standard ensemble for MC is {\em{the canonical ensemble}} due to the natural appearance of temperature in the Boltzmann factor, in which the total number $N$ of particles, the volume $V$, and the equilibrium temperature $T$ of the system are held fixed. Here, {\em{standard}} means only to pursue the relations existing in theory, and without applying any computational trick to obtain the desired result. By a similar reasoning, the standard ensemble for MD is however {\em{the microcanonical ensemble}}, in which, $N$, $V$, and the total energy $E$ are held fixed. In other words, by solely writing down the classical equations of motion in a computer language, a fixed total energy for the system of interest is then yielded. However, to simulate more realistic situations, one has to keep temperature or pressure of the system fixed, for which other statistical ensembles (say the canonical ensemble) must accordingly be used.
\subsection{Interparticle interaction}
\begin{figure}[H]
	\centering
	\fbox{\rule[0cm]{0cm}{0cm} \rule[0cm]{0cm}{0cm}
	\includegraphics[scale=0.35]{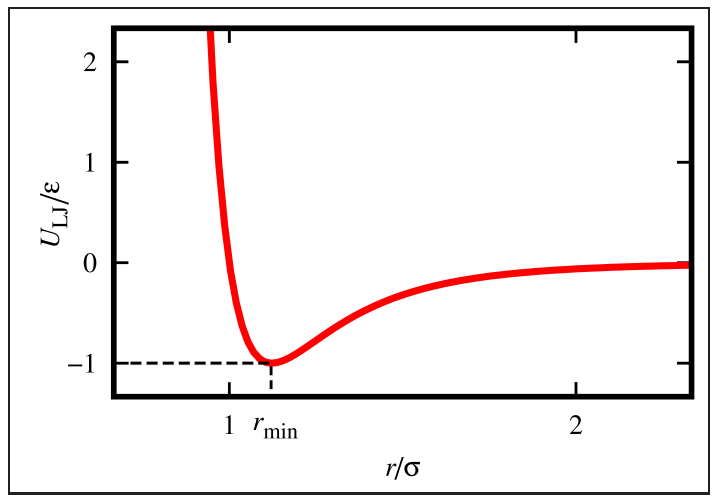}}
	\caption{\label{fig:1}
		The Lennard-Jones pairwise potential ($U_{LJ}$) as a function of the distance $r$ between two interacting particles---rendered in Gnuplot (version 5.2)~\cite{24}. $r_{min}$, corresponding with the minimum value of the potential, takes place at $2^{1/6}\sigma$.}
\end{figure}

One of the well-known interparticle potentials used in MD simulations is the Lennard-Jones (L-J) pair interaction, which is the most prevalent among other types of pairwise potentials (e.g., Morse~\cite{21}, Yukawa~\cite{22}, Coulomb, gravitational, and Buckingham~\cite{23}), and is considered as the archetype model for simple, yet-realistic interacting systems. In the L-J potential, the pairwise interaction is indeed of the van der Waals form, namely $r^{-6}$ with $r$ the interparticle distance, which describes attraction at long distances. This term could in fact be derived solely by using quantum mechanics. Adding repulsion at short distances via $r^{-12}$---according to the Pauli repulsion---then results in the L-J potential:
\begin{equation}
\label{eq:lj}
U_{LJ}(r)=-4\epsilon\bigg[\left(\frac{\sigma}{r}\right)^{6}-\left(\frac{\sigma}{r}\right)^{12}\bigg],
\end{equation}
where $r=|{\bf{r}}_{1}-{\bf{r}}_{2}|$, ${\bf{r}}_{1}$ and ${\bf{r}}_{2}$ denote respectively the position vectors of particles 1 and 2, $\epsilon$ is the depth of the potential well (also referred to as the dispersion energy), and $\sigma$---often referred to as the particle size---is the distance at which $U_{LJ}$ vanishes (bold characters denote vectors). $r_{min}=2^{1/6}\sigma$ also corresponds with the minimum of the potential, as illustrated in Fig.~\ref{fig:1} 
\subsection{Long- and short-range interactions: The cutoff radius}
\begin{figure}[H]
	\centering
	\fbox{\rule[0cm]{0cm}{0cm} \rule[0cm]{0cm}{0cm}
	\includegraphics[scale=0.5]{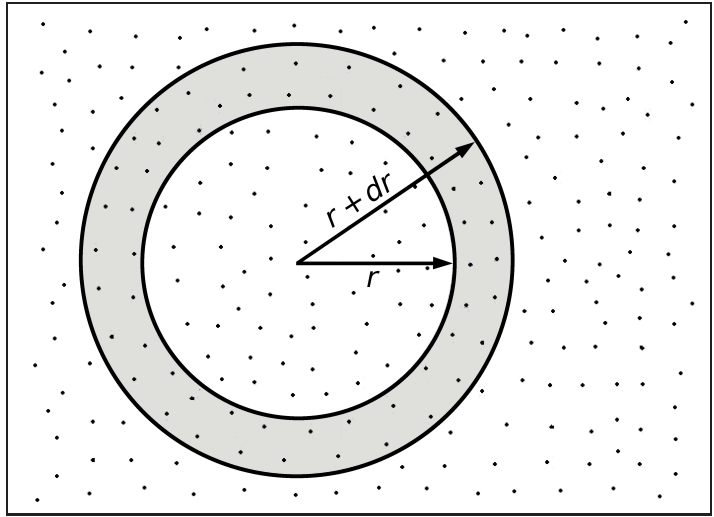}}
	\caption{\label{fig:2}
		A universe of particles with a uniform distribution, and the two spheres with radii $r$ and $r+dr$ encompassing parts of the particles---rendered in GIMP (version 2.8.22)~\cite{25}.}
\end{figure}

Assume a universe of particles with a uniform distribution (Fig.~\ref{fig:2}) interacting via the pairwise potential $U$, which is proportional to $r^{-\alpha}$ ($\alpha>0$).
The question is that could it be reliable ignoring the collective effect of particles outside the sphere of radius $r_c$ on every point inside that sphere, and taking into account only the interactions of particles within that sphere, in order to merely reduce the computational cost? 

The answer fundamentally depends on the value of $\alpha$. Assuming $\rho$ to be the particle density in the universe of Fig.~\ref{fig:2}, then
\begin{equation*}
\Delta U(r)=\frac{\rho\hspace{1mm}4\pi r^{2}dr}{r^{\alpha}}\propto r^{2-\alpha}.
\end{equation*}
Now, if $\alpha<2$, then $\lim_{r\longrightarrow\infty}\Delta U=\infty$, and the interaction [i.e., $U(r)$] is said to be long-range. A clear illustration of this kind is the gravitational force for which $\alpha=1$, and therefore, $\Delta U(r)\propto r$ meaning that the farther the planet is from say Earth, the larger its potential energy effect on Earth. Accordingly, gravitational potential energy contributions of very distant celestial objects say at the edge of the galaxy, to our planet are considerably larger than that of the Sun, while the net contribution of their forces could be zero due to vector summation. Therefore, it is not reasonable to ignore particles outside the sphere if the pair interaction is gravitational. 

In contrast, if $\alpha>2$, then $\lim_{r\longrightarrow\infty}\Delta U=0$, and the interaction is said to be short-range, such as the L-J potential. As a result, we can eliminate the particles outside a so-called cutoff radius ($r_c$), without loss of generality or reliability of calculations, in order to reduce the computational cost.
\subsection{Choosing appropriate system of units: Making physical observables dimensionless}
A given system of units, as such, is not more advantageous than other ones, and we, in theory, use a specific system of units only to speak in the same scientific language. However, in computer simulations particularly MD, choosing appropriate systems of units is of foremost importance. We, for example, cannot use the meter length scale in simulating nanoscale systems, or, it is not feasible using electron Volt energy scale when the process of interest is nuclear fusion in the Sun. In either case, one would get very large or very small numbers, and it is accordingly needed a huge amount of memory to save them, regardless of the rise of round-off errors.

Consider the simulation of a system composed of argon atoms interacting via the L-J potential. The mass $m_A$ of an argon atom, and the related L-J parameters, namely $\epsilon_{A}$ and $\sigma_{A}$, are tabulated in Table~\ref{tab:1}.
\begin{table}[H]
	\small
	\caption{The mass ($m_A$) of an argon atom, as well as the L-J parameters $\epsilon_{A}$, and $\sigma_{A}$ associated with a pair of argon atoms~\cite{26}.}
	\label{tab:1}
	\begin{tabular}{ccc}
		\hline
		\raisebox{-1ex}{$m_A$ (kg)}
		& \raisebox{-0.8ex}{$\epsilon_{A}$ (J)}
		& \raisebox{-0.8ex}{$\sigma_{A}$ (m)}\\
		\hline
		\raisebox{-0.8ex} {$6.7\times 10^{-26}$} & \raisebox{-0.8ex} {$1.6\times 10^{-21}$} & \raisebox{-0.8ex} {$3.4\times 10^{-10}$}\\
		\hline
	\end{tabular}
\end{table}
As a result, $m^{*}\equiv m\big/m_{A}$, $E^{*}\equiv E\big/\epsilon_{A}$, and $r^{*}\equiv r\big/\sigma_{A}$ are respectively the dimensionless mass, energy, and distance, which form {\em{the reduced units}}. These dimensionless quantities, in fact, keep the MD simulation from generating very large or very small quantity values. Since the dimensions of mass, energy, and length are respectively M, ML$^2$T$^{-2}$, and L, the characteristic time for simulating a system of argon atoms, using values of Table~\ref{tab:1}, is then obtained as 
\begin{equation*}
\sqrt{\mathrm{\frac{ML^{2}}{ML^{2}T^{-2}}}}={\mathrm{T}}\hspace{3mm}\Longrightarrow\hspace{3mm}\tau_{A}=\sigma_{A}\sqrt{\frac{m_{A}}{\epsilon_{A}}}\cong 10^{-12}\hspace{1mm}{\mathrm{s}}=1\hspace{1mm}{\mathrm{ps}}.
\end{equation*}
The dimensionless time is also then $t^{*}\equiv t\big/\tau_A$. This specific value, namely $\tau_A$, is indeed the characteristic timestep of a gas of argon atoms for which a logical relationship between any two consecutive events holds according to the principle of causality. For timesteps considerably larger than $\tau_A$, the simulation crashes because of loss of causality, while values much smaller than $\tau_A$ increase the computational cost and make accordingly the simulation dramatically slow. Timesteps of about $10^{-3,-4}\tau_{A}=10^{-15,-16}\hspace{1mm}{\mathrm{s}}$ are optimal for simulating systems at the scales of atoms and molecules.

The characteristic velocity is also $v_{A}=\sigma_{A}\big/\tau_{A}\cong 340\hspace{1mm}{\mathrm{m\big/s}}$, which is nearly the same velocity of sound. This, in fact, is reasonable because sound is nothing but the collisions of particles which propagate through a transmission medium. The dimensionless velocity is accordingly $v^{*}\equiv v\big/v_{A}$. The characters indicated by asterisks are called {\em{derived quantities}} due to being derived by {\em{fundamental quantities}} tabulated in Table~\ref{tab:1}.

As a consequence, property values generated by the computer code (i.e., outputs) are also evidently dimensionless, and one can re-attribute physical meanings to them by pursuing the reverse procedure. If, for example, the final time of the simulation is reported $10^7$, it means that
\begin{equation*}
t^{*}=\frac{t}{\tau_{A}}=10^{7}\hspace{3mm}\Longrightarrow\hspace{3mm} t=10^{7}\hspace{1mm}\tau_{A}=10^{7}\times 10^{-16}\hspace{1mm}{\mathrm{s}}=1\hspace{1mm}{\mathrm{ns}}
\end{equation*}
\section{\label{sec:3}Simulation preliminaries}
\subsection{\label{sec:31}Initialization}
The first step in MD simulations at the level of programming is to assign initial values to each Cartesian component of atomic positions, velocities, and accelerations. One, also, has to set the total number $N$ of particles and the size $L$ of the simulation box within which particles move. Assume there exists a three-dimensional, uniform lattice at each point of which an argon atom is located, with a total number of 1000 atoms. As a result, $L$ must be $10\hspace{0.5mm}\sigma_{A}$ according to the preceding discussions. Simulating a gas of argon atoms, this value must accordingly be multiplied by the factor 2 or 3, and then $L=30\hspace{0.5mm}\sigma_{A}$.

The easiest way to generate initial positions is to use random generators, which, in most cases, does not cause a problem. A serious drawback of this method, however, is that the positions of two atoms could be very close to each other. That being so, the repulsive force between them becomes very strong, which, in turn, throws them to infinitely large distances, and the simulation would accordingly crash as well. Using random number generators with uniform distributions, however, could prevent this issue to a large extent. 

The randomness of initial velocities and accelerations also is not problematic, and we simply could set them to zero. Although random initial velocities means that the velocity distribution of particles at $t=0$ is not of the Maxwell--Boltzmann form at all, the system corrects itself after a number of timesteps.
\subsection{The center-of-mass reference frame}
We know that the temperature $T$ of a $d$-dimensional system is given by the equipartition theorem as
\begin{equation}
\label{eq:ept}
T=\frac{1}{Nk_{B}d}\Bigg\langle\sum_{i=1}^{N}m_{i}v_{i}^{2}\Bigg\rangle.
\end{equation}
where the $v_i$ values are calculated with respect to the center-of-mass (COM) reference frame. Indeed, the average temperature of say a metal rod in my hand should not, at all, be dependent on the velocity at which I walk. In other words, calculating temperature using Eq.~\ref{eq:ept} is valid only in the COM reference frame, therefore, a computer routine must be written which resets the COM velocity to zero, or equivalently, subtracts the COM velocity from the atomic velocities, every once in a while.
\subsection{Solving differential equations of motion}
If $n$ denotes the timestep index, $h$ the timestep, $v$ the velocity, $a$ the acceleration, and $\beta$ the jerk, expanding position $x$ in $h$ results in
\begin{eqnarray}
x_{n+1}=x_{n}+v_{n}h+\frac{1}{2!}a_{n}h^{2}+\frac{1}{3!}\beta_{n}h^{3}+\mathcal{O}(h^{4}),\nonumber\\
x_{n-1}=x_{n}-v_{n}h+\frac{1}{2!}a_{n}h^{2}-\frac{1}{3!}\beta_{n}h^{3}+\mathcal{O}(h^{4})\nonumber.
\end{eqnarray}
Adding and subtracting these two equations accordingly lead to
\begin{equation}
\label{eq:a1}
x_{n+1}=-x_{n-1}+2x_{n}+a_{n}h^{2}+\mathcal{O}(h^{4}),
\end{equation}
\begin{equation}
\label{eq:a2}
x_{n+1}-x_{n-1}=2v_{n}h+\mathcal{O}(h^{3})\hspace{3mm}\Longrightarrow\hspace{3mm}v_{n}=\frac{1}{2h}\left(x_{n+1}-x_{n-1}\right).
\end{equation}
The major drawback of Eq.~\ref{eq:a2} is that it brings about a large round-off error according to the fact that $x_{n+1}$ and $x_{n-1}$ are of the same order, and computer may accordingly round them to very close values, or to the same value. This is called the {\em{Verlet}} algorithm~\cite{27}. To overcome this issue, another one called the {\em{velocity Verlet}} algorithm~\cite{28} is used, which updates the atomic positions and velocities as
\begin{eqnarray}
\label{eq:vervelx}
x_{n+1}=x_{n}+v_{n}h+\frac{1}{2}a_{n}h^{2},
\end{eqnarray}
\begin{eqnarray}
\label{eq:vervelv}
v_{n+1}=v_{n}+\frac{1}{2}(a_{n}+a_{n+1})h.
\end{eqnarray}
The equivalence between these two algorithms is shown in Appendix \ref{sec:app1}. The {\em{Beeman's method}}~\cite{29} is also another alternative, in which positions and velocities are updated according to
\begin{equation*}
x_{n+1}=x_{n}+v_{n}h+\frac{1}{6}(4a_{n}-a_{n-1})h^{2},\\
\end{equation*}
\begin{equation*}
v_{n+1}=v_{n}+\frac{1}{6}(2a_{n+1}+5a_{n}-a_{n-1})h.
\end{equation*}
Almost all MD codes apply one of these two algorithms (velocity Verlet or Beeman). Here, we exploit the former to solve Newton's equations of motion. The key point, at this stage, is that the accelerations must be updated right after updating the positions and before computing the velocities. This is based on the fact that the index $n+1$ appears in both sides of Eq.~\ref{eq:vervelv}, while we have not yet the acceleration of the next step, namely $a_{n+1}$, in hand. Accordingly, (i) we first calculate positions using Eq.~\ref{eq:vervelx}; (ii) then, $v'=v_{n}+\frac{1}{2}a_{n}h$; (iii) then accelerations are calculated---because the acceleration due to the L-J potential is a function of position, calculating it after $x_{n+1}$ yields $a_{n+1}$; and finally (iv) $v_{n+1}=v'+\frac{1}{2}a_{n+1}h$.
\subsection{Calculating the forces}
From the L-J potential (i.e., Eq.~\ref{eq:lj}), the pairwise force along the $x$ direction is obtained as follows:
\begin{equation*}
U_{ij}=4\Bigg(\frac{1}{r_{ij}^{12}}-\frac{1}{r_{ij}^{6}}\Bigg)\hspace{3mm}\Longrightarrow\hspace{3mm}(f_{ij})_{x}=-\frac{\partial U_{ij}}{\partial r_{ij}}\frac{\partial r_{ij}}{\partial x}=4\Bigg(\frac{12}{r_{ij}^{12}}-\frac{6}{r_{ij}^{6}}\Bigg)\frac{x_{ij}}{r_{ij}^{2}},
\end{equation*}
where $\sigma$ and $\epsilon$ has been set to unity. That we factored out $r_{ij}^{-2}$ in the right-hand side is of foremost importance. By this seemingly frivolous simplification, $r_{ij}^{-6}$ and $r_{ij}^{-12}$ are calculated once for both the potential energy and the force; otherwise, one has to compute $r_{ij}^{-6}$ and $r_{ij}^{-12}$ for the potential, and $r_{ij}^{-8}$ and $r_{ij}^{-14}$ for the force, which dramatically increases the computational cost. Nearly all the computational cost arises from force calculation because of being inherently pairwise, in contrast to the Verlet algorithm which is essentially single-particle. More precisely, the Verlet routine is of order $N$ while the force (acceleration) routine is of order $N^2$. Therefore, for a large number $N$ of particles, the cost of the Verlet routine is dominated by that of the force. A way to speed up the simulation is then calculating the potential and the force simultaneously, as mentioned before. This cost reduction or speed up could further be increased using Newton's third law, which reduces the order $N^2$ to $N(N-1)\big/2$. In other words, instead of calculating the entire force matrix, only an upper triangular (without the diagonal elements) is calculated, and it is therefore evident that ${\bf{F}}_{ij}=-{\bf{F}}_{ji}$ gives the full matrix of interest.
\subsection{Periodic boundary conditions}
In MD simulations, periodic boundary conditions (PBC) are used for two purposes: (I) to keep fixed the number of particles within the simulation box (the primary cell) of the system; and (II) to eliminate particles out of the cutoff radius in order to reduce the computational cost.
\subsubsection{Purpose I}
\begin{figure}[H]
	\centering
	\fbox{\rule[0cm]{0cm}{0cm} \rule[0cm]{0cm}{0cm}
	\subfigure[]{\label{subfig:3(a)}
		\includegraphics[scale=0.1]{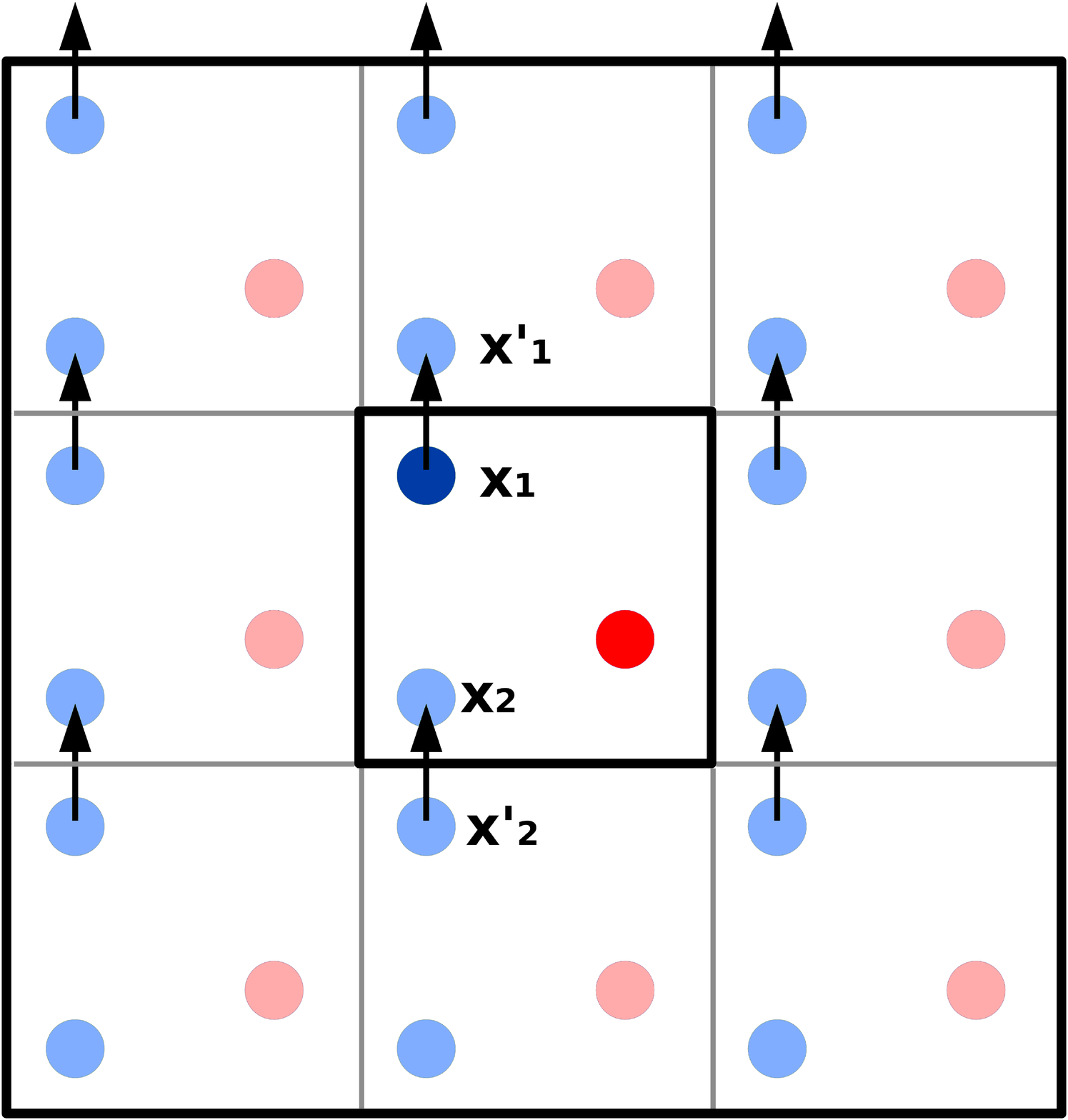}}
	\subfigure[]{\label{subfig:3(b)}
		\includegraphics[scale=0.25]{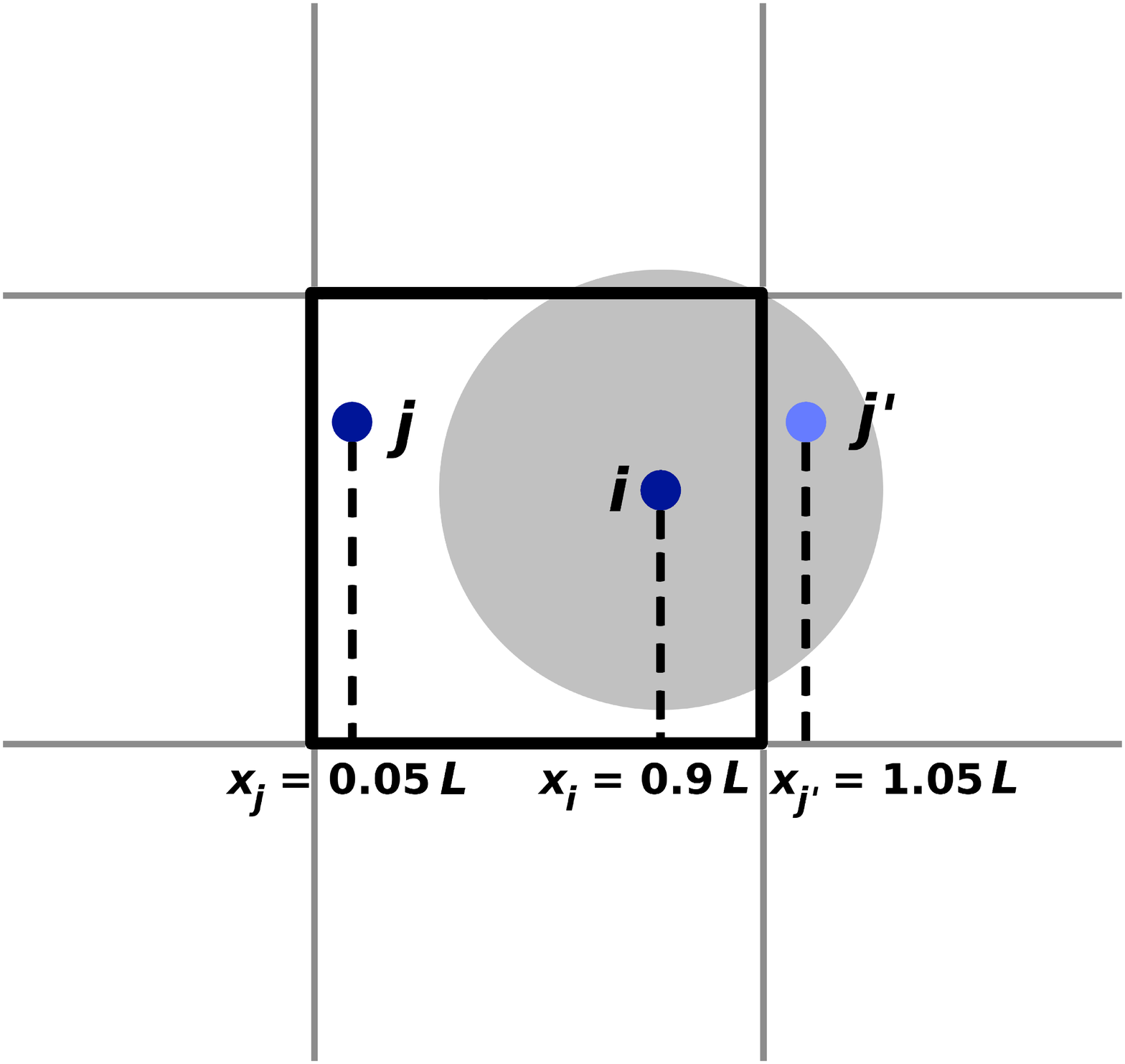}}}
	\caption{\label{fig:3}
		Schematic of PBC in two dimensions. (a) The simulation box at the center (i.e., the main system, or the primary cell) contains two particles shown in dark red and dark blue. The neighboring boxes, however, indicate the image systems composed of image particles shown in light red and light blue. As the blue particle leaves the box by moving from ${\mathrm{{\bf{X}}_{1}}}$ to ${\mathrm{{\bf{X'}}_{1}}}$, its image simultaneously enters from the opposite side from ${\mathrm{{\bf{X'}}_{2}}}$ to ${\mathrm{{\bf{X}}_{2}}}$, and this happens for all image cells. (b) The simulation box with size $L$ and the cutoff sphere of radius $r_{c}=L\big/2$ shown by the gray area. The system is composed of two particles i and j, however, i only interacts with the image of j, namely j'.}
\end{figure}

Keeping fixed the total number of particles is done simply in a way that once a particle leaves the box, a same one enters from the opposite side [Fig.~\ref{subfig:3(a)}]. This is as though the related computer routine only shifts the position of the outgoing particle so as to enter it from the opposite side of the box.
\subsubsection{Purpose II}
PBC are also necessary when we use short-range potentials, and are going to eliminate particles outside the cutoff radius. We know that because of PBC, particle $i$ within the primary cell interacts not only with other particles within the primary cell, but also with those within image cells. Assume the box size to be $L$ and the cutoff radius $r_{c}=L\big/2$. Therefore, as seen in Fig.~\ref{subfig:3(b)}, particle $i$ only interacts with the image of $j$, namely $j'$, because
\begin{eqnarray*}
	R_{ij}=x_{i}-x_{j}=0.9\hspace{1mm}L-0.05\hspace{1mm}L=0.85\hspace{1mm}L>r_{c},
\end{eqnarray*}
while
\begin{eqnarray*}
	R_{ij'}=x_{j'}-x_{i}=1.05\hspace{1mm}L-0.9\hspace{1mm}L=0.15\hspace{1mm}L<r_{c}.
\end{eqnarray*}
In such a case, $R_{ij}$ must be replaced by $R_{ij'}$. This replacement, at the programming level, corresponds with the assignment statement \texttt{Rij=Rij-sign(1.,Rij)}, where \texttt{sign} is the sign function.

\begin{figure}[H]
	\centering
	\fbox{\rule[0cm]{0cm}{0cm} \rule[0cm]{0cm}{0cm}
	\includegraphics[scale=0.65]{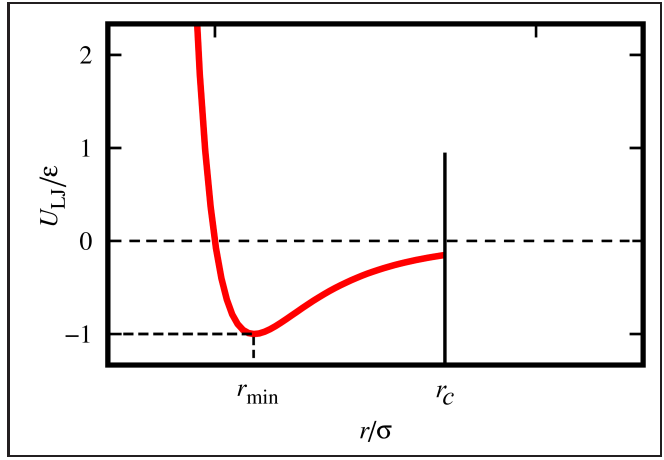}}
	\caption{\label{fig:4}
		The L-J potential (Eq.~\ref{eq:lj}) truncated at $r_{c}=2.50\hspace{0.5mm}\sigma$. That $U_{LJ}(r_c)\neq 0$ brings about an issue in a way that the interaction force exerted on a particle suddenly vanishes at the instance it is leaving the cutoff radius.}
\end{figure}

For the L-J potential, $r_c$ is usually set to $2.50\hspace{1mm}\sigma$. The issue associated with the cutoff radius is that $U_{LJ}$ is not differentiable at $r=r_c$ (Fig.~\ref{fig:4}), therefore, the force at this point cannot be calculated.
As is seen, $U_{LJ}(r_c)\neq 0$, and this, in turn, poses a problem in a way that the particle experiences an abrupt freedom at the instance it is leaving the cutoff sphere. The simplest way to overcome such a discontinuity is to so-called bend the potential, which is done by shifting $U_{LJ}$, namely $U_{LJ}\longrightarrow U_{LJ}-U_{LJ}(r_c)$. 

However, there still exist a couple of issues related to such a modification, which are (1) fluctuation in energy of the particle, which decreases [$E=mv^{2}/2+U_{LJ}-U_{LJ}(r_c)$] when entering the cutoff radius and increases ($E=mv^{2}/2+U_{LJ}$) when leaving it; and (2) $U_{LJ}$ is not still differentiable at $r_c$. The transformation $U_{LJ}\longrightarrow U_{tot}=U_{LJ}-U_{LJ}(r_c)-rU_{LJ}'(r_c)$ indeed fixes the problems. As an illustration, for the $x$ component of the force at $r_c$, we have
\begin{equation*}
f_{x}(r_c)=\frac{dU_{tot}}{dx}\Bigg|_{r_c}=\frac{dU_{tot}}{dr}\Bigg|_{r_c}\frac{dr}{dx}=\frac{dU_{LJ}}{dr}\Bigg|_{r_c}-0-U_{LJ}'(r_c)=0,	
\end{equation*}
where we have used
\begin{equation*}
\frac{dU_{tot}}{dr}=24\epsilon\bigg(-2\frac{\sigma^{12}}{r^{13}}+\frac{\sigma^{6}}{r^{7}}\bigg)-U_{LJ}'(r_c),
\end{equation*}
and, $U_{LJ}'=dU_{LJ}\big/dr$.
\subsection{Thermodynamic equilibrium}
\begin{figure}[H]
	\centering
	\fbox{\rule[0cm]{0cm}{0cm} \rule[0cm]{0cm}{0cm}
	\subfigure[]{\label{subfig:5(a)}
		\includegraphics[scale=0.309,angle=-90]{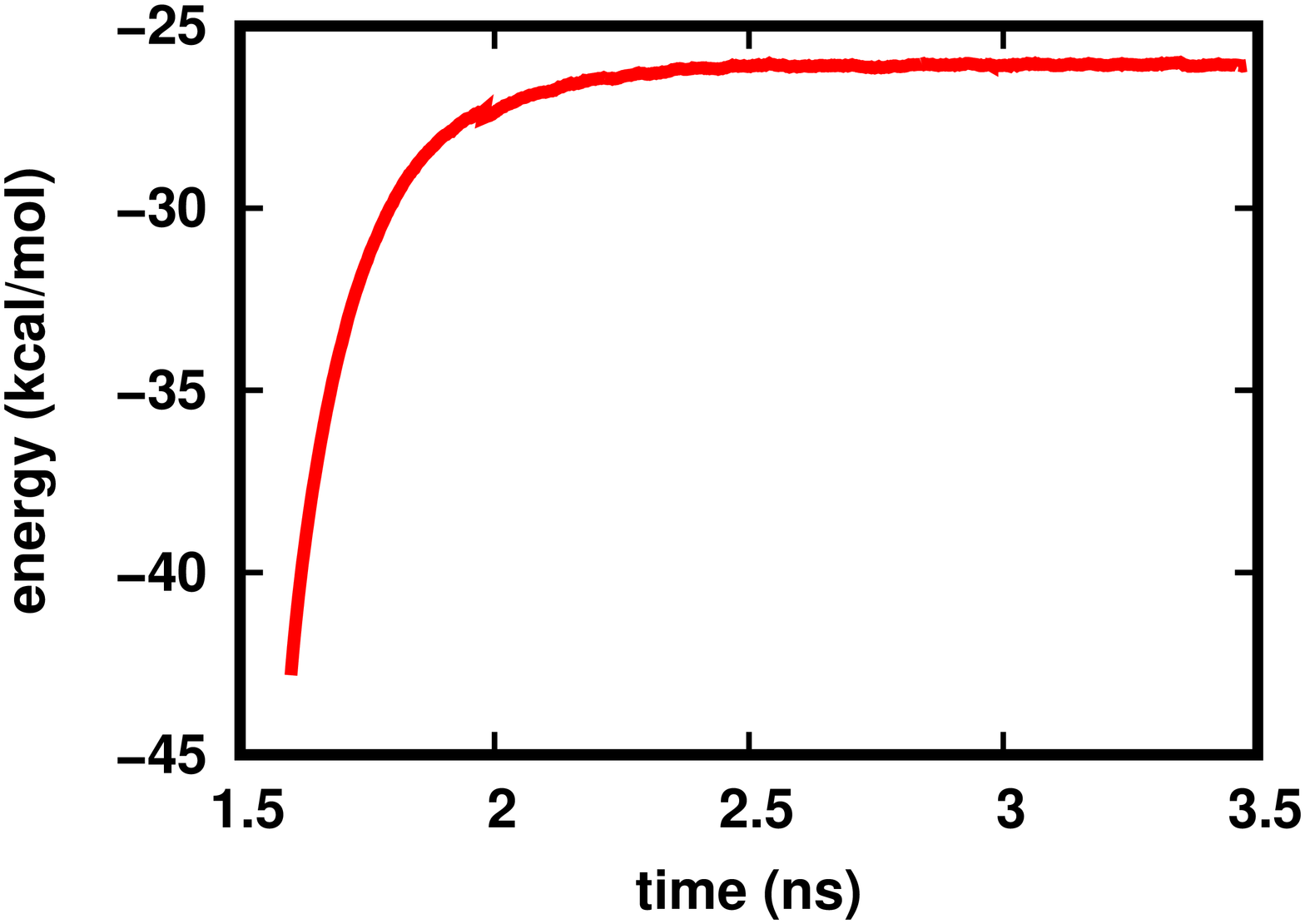}}
	\subfigure[]{\label{subfig:5(b)}
		\includegraphics[scale=0.309,angle=-90]{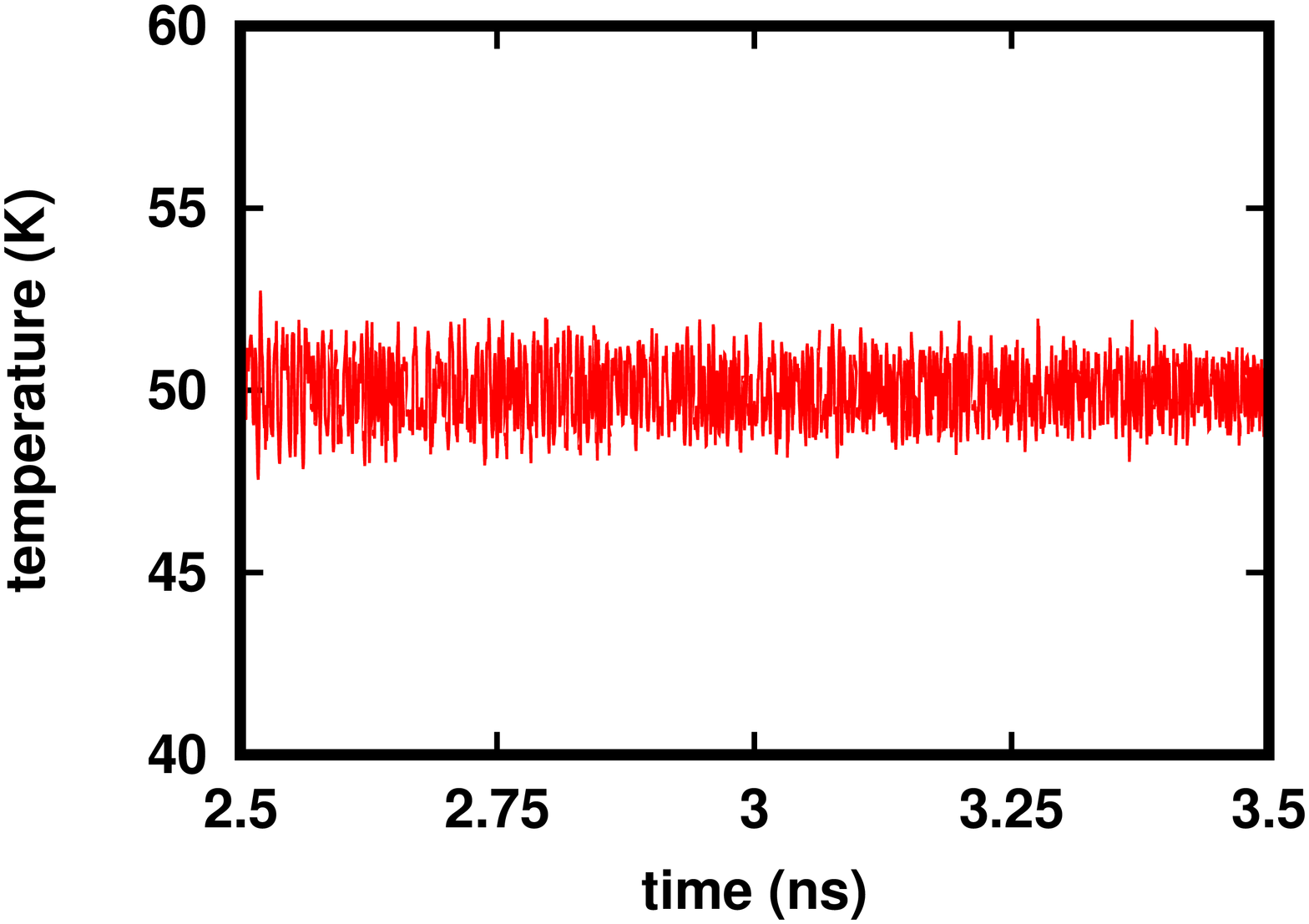}}}
	\caption{\label{fig:5}
		(a) Time dependence of the total energy of a system during an MD simulation. Over 2.5 to 3.5 ns, the energy converges meaning that the system approaches equilibrium from 2.5 ns on. In contrast, the system is entirely out of equilibrium over 1.5 to 2 ns due to a sharp variation. (b) Time dependence of the instantaneous temperature of the system discussed in (a), showing a perfect converging pattern over 2.5 to 3.5 ns of the simulation, with the equilibrium temperature of 50 K.}
\end{figure}

From when on, is the simulation reliable for calculating statistical averages? The answer is that when the system approaches some state of thermodynamic equilibrium. Indeed, starting from random initial conditions (random positions, and velocities) definitely means the system is initially far from equilibrium, and this is while all thermodynamic properties (say temperature) are defined only at equilibrium. One way to check if the system has approached a state of thermodynamic equilibrium is to plot the time dependence of its total energy, as illustrated in Fig.~\ref{subfig:5(a)} for an MD simulation.
As is seen, the system is out of equilibrium over the first 2.5 ns because of a sharp variation and lack of convergence, in contrast to the last 1 ns with a perfect convergence, meaning that the system has reached a state of equilibrium. Time dependence of the instantaneous temperature is also very useful for checking if the system is in equilibrium. Fig.~\ref{subfig:5(b)} illustrates such a diagram associated with the last 1 ns of Fig.~\ref{subfig:5(a)}, showing that the system is in thermodynamic equilibrium, with an average temperature of 50 K.
\subsection{Degrees of freedom}
In MD simulations, a $d$-dimensional system composed of $N$ particles has $(N-1)d$ degrees of freedom (DOF) instead of $Nd$. Indeed, based on the constraint that the COM velocity ($v_{cm}$) must be zero, there then exist $(N-1)$ DOF for each Cartesian component of atomic velocities, according to which there is at least one velocity being dependent on the rest:
\begin{eqnarray*}
	v_{cm}=\sum_{i=1}^{N}v_{i}=0\hspace{3mm}\Longrightarrow\hspace{3mm}v_{1}+v_{2}+...+v_{N}=0\hspace{3mm}\Longrightarrow\hspace{3mm} v_{1}=-(v_{2}+...+v_{N}).
\end{eqnarray*}
In the thermodynamic limit, $N-1\simeq N$, however, for limited numbers of particles as in all MD simulations, say $N=100$, the associated error would then be 1\%, which is not negligible at all.

It is also possible to have further DOF, such as simulating water molecules with rotational and vibrational DOF, each of which gives $k_{B}T\big/2$ according to the equipartition theorem. One can use this theorem only in say $x$ direction to estimate the temperature of the system, without the need to know all components of the velocities. One way to test the reliability of MD simulations, in fact, is that the temperature values obtained from different DOF, or along different spatial directions, must be the same. 
\subsection{Velocity autocorrelation function}
\begin{figure}[H]
	\centering
	\fbox{\rule[0cm]{0cm}{0cm} \rule[0cm]{0cm}{0cm}
	\includegraphics[scale=0.32,angle=-90]{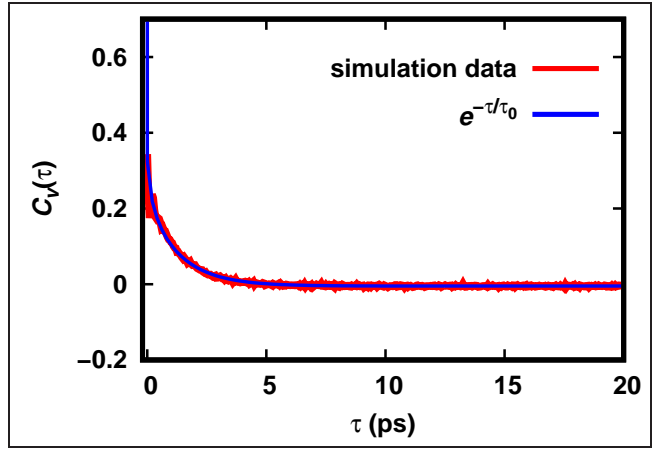}}
	\caption{\label{fig:6}
		Velocity autocorrelation function for an MD simulation. Fitting the exponential function $e^{-\tau/\tau_{0}}$ (blue) to the obtained curve (red) gives $\tau_{0}$ as the equilibrium time of the system.}
\end{figure}

Using the velocity autocorrelation function [$C_{v}(\tau)$], one could have an estimation of the equilibrium time of the system. Indeed, $C_{v}(\tau)$ measures how velocity at $t+\tau$ is correlated with its value at $t$, giving therefore an estimation of the time during which the system loses its memory of previous atomic velocities. By definition,
\begin{equation}
\label{eq:ct}
C_{v}(\tau)=\frac{\langle v_{i}(t)v_{i}(t+\tau)\rangle-\langle v_{i}(t)v_{i}(t-\tau)\rangle}{\sum_{i=1}^{N}\frac{1}{t_{0}}\int_{0}^{t_0}v_{i}^{2}(t)dt},
\end{equation}
where the denominator is equal to $(N-1)dT$ for a $d$-dimensional system of $N$ particles at the equilibrium temperature $T$. $\langle v(t)\rangle$ is always zero based on the following reasoning. If the system is large enough, after an infinite time, and within a uniform and isotropic universe, the particles would accordingly move along all directions and the average velocity is then zero. However, the simulation time is very limited, but $\langle v(t)\rangle$ is still zero because
\begin{equation*}
\langle v(t)\rangle=\sum_{i=1}^{N}\int v_{i}(t)dt=\int\sum_{i=1}^{N}v_{i}(t)dt=\int v_{cm}dt=0.
\end{equation*}
As a result, the second term in the numerator of Eq.~\ref{eq:ct} vanishes, and then
\begin{equation*}
C_{v}(\tau)=\frac{\langle v_{i}(t)v_{i}(t+\tau)\rangle}{(N-1)dT}.
\end{equation*}
Fig.~\ref{fig:6} illustrates a typical $C_{v}(\tau)$ for an MD simulation.
As is seen, fitting an exponential function of the form $e^{-\tau/\tau_{0}}$ to the obtained simulation data results in the value of $\tau_0$ as an estimation of the equilibrium time of the system.
\subsection{Calculating the pressure}
There are three ways to compute the pressure $P$ of the system in MD simulations, which are explained as follows.
\subsubsection{Method 1}
Define a plane with area $A$; compute the total number (say $N_0$) of particles crossing the plane during the time interval $\Delta t$; compute their associated linear momenta; and then
\begin{eqnarray*}
	\Delta p=\sum_{i=1}^{N_0} 2m_{i}v_{i}=F\Delta t\hspace{3mm}\Longrightarrow\hspace{3mm}F=\frac{\sum_{i=1}^{N_0} 2m_{i}v_{i}}{\Delta t}\hspace{3mm}\Longrightarrow\hspace{3mm}P=\frac{F}{A}=\frac{\sum_{i=1}^{N_0} 2m_{i}v_{i}}{A\Delta t},
\end{eqnarray*}
where $\Delta p$ is the total momentum transferred to the plane, $F$ is the associated total force, and $m_i$ and $v_i$ are respectively the mass and velocity of particle $i$.
\subsubsection{Method 2}
Import one particle with the opposite velocity for each one leaving the simulation box. Therefore
\begin{eqnarray*}
	\Delta p=F\Delta t=p_{out}-p_{in}\hspace{3mm}\Longrightarrow\hspace{3mm}F=\frac{\sum_{i}(mv_{i,\hspace{1mm}out}-mv_{i,\hspace{1mm}in})}{\Delta t}\hspace{3mm}\Longrightarrow\hspace{3mm}P=\frac{F}{A}=\frac{\sum_{i}(mv_{i,\hspace{1mm}out}-mv_{i,\hspace{1mm}in})}{A\Delta t},
\end{eqnarray*}
where $p_{in}$ ($p_{out}$) is the linear momentum of the particle entering (leaving) the simulation box. It should be noted that $\Delta t$ must be large enough here to have a relatively fair approximation of the pressure.
\subsubsection{Method 3}
Use the kinetic theory of gases. For an ideal gas~\cite{30}, $P=(N/V)k_{B}T$. However, in a more general case, if the van der Waals pair interaction is added to that ideal, $d$-dimensional system, then
\begin{eqnarray}
\label{eq:p}
P=\frac{N}{V}k_{B}T+\frac{1}{Vd}\Bigg\langle \sum_{i<j}\mathrm{\bf{F}}(\mathrm{\bf{r}}_{ij}).\mathrm{\bf{r}}_{ij}\Bigg\rangle=\frac{N}{V}k_{B}T-\frac{1}{Vd}\Bigg\langle \sum_{i<j}\frac{\partial U_{ij}}{\partial r_{ij}}r_{ij}\Bigg\rangle,
\end{eqnarray}
which is the virial equation of state. The superiority of this method, compared to the previous, is that Eq.~\ref{eq:p} uses all the particles to compute the pressure, also gives the instantaneous pressure of the system, without any need to wait by $\Delta t$.

The virial of the system, namely $\sum_{i}q_{i}{\dot{p}_{i}}=\sum_{i}{\mathrm{\bf{F}}}_{i}.{\mathrm{\bf{r}}}_{i}$, could have also been used to compute the pressure according to
\begin{equation}
\label{eq:p2}
P=\frac{N}{V}k_{B}T+\frac{1}{Vd}\Bigg\langle \sum_{i}\mathrm{\bf{F}}_{i}.\mathrm{\bf{r}}_{i}\Bigg\rangle.
\end{equation}
However, due to PBC, for a particle leaving the box, and for its image entering from the opposite side, the two single-particle forces may be very similar, while a large jump in the single-particle position definitely takes place, leading as well to an abrupt change in the virial and then in the estimated pressure of the system. Indeed, it was this issue that led us to use double-particle forces and positions, as implemented in Eq.~\ref{eq:p}. However, one could show the equivalence between Eqs.~\ref{eq:p} and~\ref{eq:p2} (see Appendix~\ref{sec:app2}).
\subsection{Thermostat: Controlling the temperature}
As mentioned before, the standard ensemble for MD is the microcanonical, in that the macrostate of the system is defined through the fixed numbers $N,V$, and $E$ [or, more realistically, a fixed energy range ($E, E+\Delta$)]. The basic problem is then to determine the total number $\Omega(N,V,E)$ [or, $\Omega(N,V,E;\Delta)$] of distinct, accessible microstates, from which complete thermodynamics of the system could be derived straightforwardly. For most physical systems, however, determining $\Omega$ is quite formidable if not intractable. More importantly, the concept of a fixed energy, or even a fixed energy range, for a real-world system is not satisfactory at all, based on the fact that the total energy $E$ of a system is hardly ever measured; and therefore, it is not possible to keep its value fixed in the lab. 

An alternative to fixed energy is fixed temperature, which not only is measured directly in the lab using thermometers, but also is controllable and could be kept fixed by keeping the system in contact with an appropriate thermal bath. Such a so-called generalization is referred to as the canonical ensemble, in that the macrostate of the system is defined by the fixed numbers $N,V$, and $T$. This is indeed one of the most prevalent ensembles used for MD simulations.

However, keeping fixed the temperature of the system in MD simulations needs applying an extra, merely-computational trick, according to the fact that temperature, in MD simulations, becomes very quickly out of control, and takes very large values far from the defined target equilibrium temperature. This trick, at the programming level, is called {\em{thermostat}}, and makes the instantaneous temperature of the system to fluctuate around the target one. Although this routine works according to the equipartition theorem (Eq.~\ref{eq:ept}), its presence in MD codes has no theoretical basis and is merely a matter of technique. From Eq.~\ref{eq:ept}, the temperature $T$ of the system is directly correlated with atomic velocities---it is of foremost importance to mention that Eq.~\ref{eq:ept}, in fact, gives the instantaneous temperature, or equivalently, the average temperature of the system at one timestep. As a result, to control the instantaneous temperature, one has to place constraint on atomic velocities. To do so, assume that the target temperature is $T_{0}$, which is not equal to instantaneous temperature $T$. Therefore, it is natural to assume that the atomic velocity $v_i$ times a dimensionless factor (say $\beta$) eventually leads exactly to $T_0$:
\begin{equation*}
\label{eq:ep2}
\frac{1}{Nk_{B}d}\Bigg\langle\sum_{i=1}^{N}m_{i}(\beta v_{i})^{2}\Bigg\rangle=T_{0}\hspace{2mm}\Longrightarrow\hspace{2mm}\beta^{2}T=T_{0}\hspace{2mm}\Longrightarrow\hspace{2mm}\beta=\Bigg(\frac{T_{0}}{T}\Bigg)^{1/2}.
\end{equation*}
As a result, the assignment statement $v_{i}=\beta v_{i}$ at the programming level, makes the average of $T$ to be very close to $T_0$. This kind of controlling the temperature is called {\em{rescaling}}. This is indeed due to rescaling the instantaneous atomic velocities that the time dependence of the thermodynamic quantities of interest (say energy, or temperature) show fluctuating patterns: every decrease after an increase in say the diagrams of Fig.~\ref{fig:5} is accordingly the effect of multiplying the atomic velocities by $\beta$.

This method is excellent in terms of computational cost, however, it may disrupt the Maxwell--Boltzmann velocity distribution of particles. To minimize such an adverse effect, one has accordingly to apply velocity rescaling at every timestep.
\section{\label{sec:4}Programming}
\begin{figure}[H]
	\centering
	\fbox{\rule[0cm]{0cm}{0cm} \rule[0cm]{0cm}{0cm}
	\includegraphics[scale=0.1]{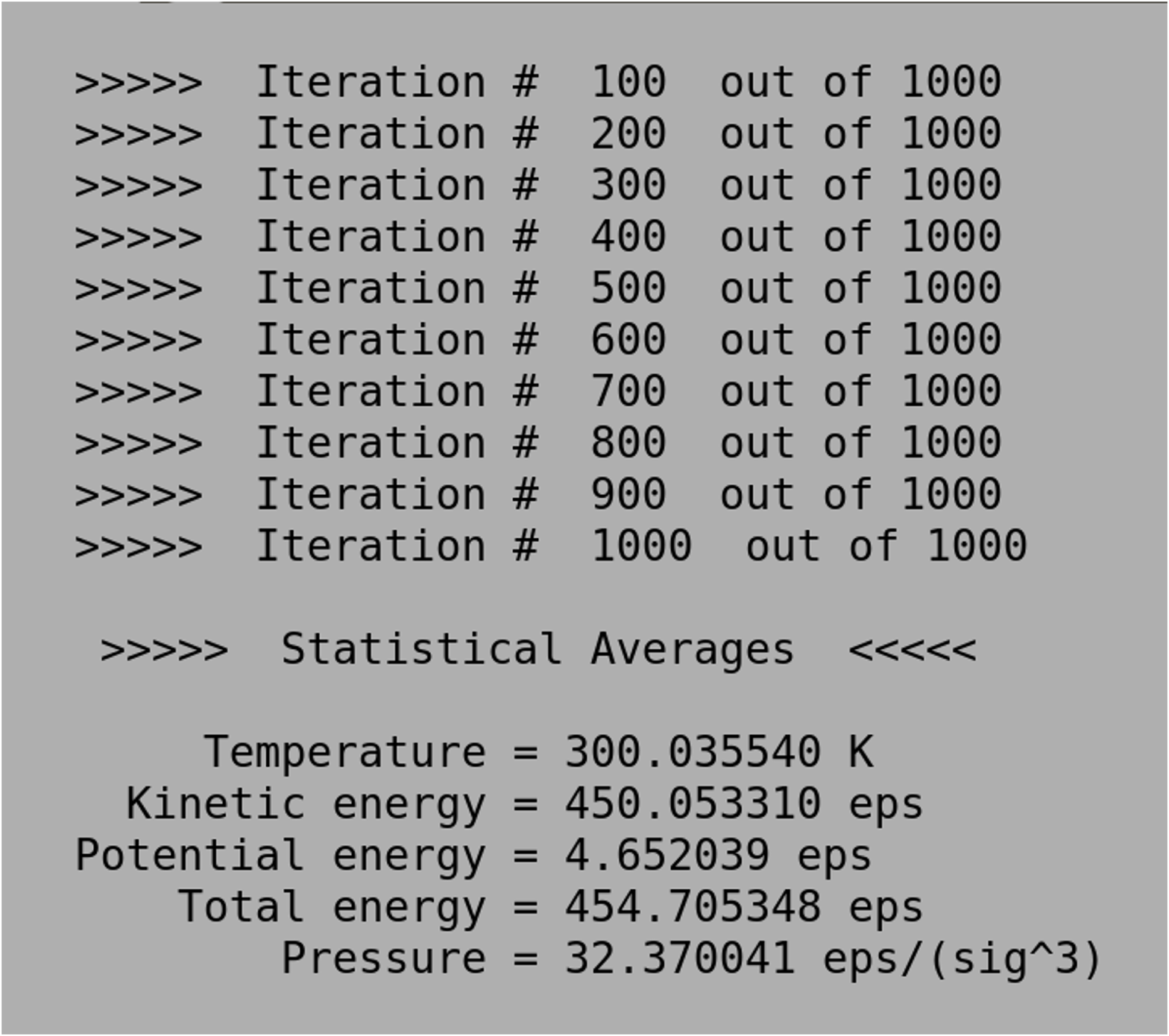}}
	\caption{\label{fig:7}
		A typical true output of \texttt{code1.py}. The first ten lines are solely to inform the user that the code is being run. {\texttt{eps}} and {\texttt{sig}} respectively stand for $\epsilon$ and $\sigma$ in Eq.~\ref{eq:lj}.}
\end{figure}

We have written the MD code---simulating a gas of classical, Lennard-Jones particles---in two forms in Python language (version 3.6.7). {\texttt{code1.py}} was written using {\em{top-down design}}, in which one starts with a large task and breaks it down to smaller pieces of program or subtasks. This is the same usual programming style used by beginners. {\texttt{code2.py}}, on the other hand, was written in a more advanced way by making use of external procedures that code each subtask as a function, each of which can be tested independently as well. It should also be noted that because the random generators applied here generate different sets of random numbers every time they are called, the outputs of any two independent runs are not exactly the same, and there is always a negligible difference. In contrast, if the difference is colossal, it then means that the issue described in Sec.~\ref{sec:31} regarding very close random initial positions of one or more pairs of atoms is encountered. Therefore, it is highly recommended to run each code several times to obtain the true set of output values (see Fig.~\ref{fig:7} for a typical true output). The notation used in these codes is also as follows.
\subsection{Notation}
\begin{tabular}{lp{0.8\textwidth}}
	{\bf{Symbol}}&{\bf{Description}}\\
	{\texttt{\_\_d\_\_}} & Dimension of the simulation box (the primary cell)\\
	{\texttt{NS}}& Total number of timesteps\\
	{\texttt{N}}& Number of particles\\
	{\texttt{dt}}& Time step\\
	{\texttt{T\_0}}& Target temperature\\
	{\texttt{sig}}& $\sigma$\\
	{\texttt{eps}}& $\epsilon$\\
	{\texttt{r\_ctf}}& Cutoff radius\\
	{\texttt{u\_at\_ctf}}&Value of the potential at the cutoff\\
	{\texttt{du\_at\_ctf}}&Value of the derivation of the potential at the cutoff\\
	{\texttt{bs}} & Box size\\
	{\texttt{vol}} & Box volume\\
	{\texttt{rho}} & Particle density\\
	{\texttt{ign}} & The first "ign" number of steps to be ignored for reliable statistical averaging, during which the system approaches a state of thermal equilibrium starting from an entirely out-of-equilibrium initial state due to random initial conditions\\
	{\texttt{pos}}&Atomic positions\\
	{\texttt{vel}}&Atomic velocities\\
	{\texttt{acc}}&Atomic accelerations\\
	{\texttt{com}}&Center of mass position\\
	{\texttt{P\_S}} & Sum of pressures\\
	{\texttt{k\_S}} & Sum of kinetic energies\\
	{\texttt{p\_S}} & Sum of potential energies\\
	{\texttt{T\_S}} & Sum of temperatures\\
	{\texttt{f\_tpz}} & Opens the file containing temperature, pressure, and compressiblility factor ({\texttt{tpz.out}})\\
	{\texttt{f\_kpe}} & Opens the file containing kinetic, potential, and total energies ({\texttt{kpe.out}})\\
	{\texttt{f\_xyz}} & Opens the file containing position coordinates ({\texttt{pos.xyz}})\\
	{\texttt{f\_AVG}} & Opens the file containing statistical averages ({\texttt{means.out}})\\
	{\texttt{sign}} & Sign function\\
	{\texttt{R}} & $r_{ij}$ scaled to box size\\
	{\texttt{r}} & $r_{ij}$ in real units\\
	{\texttt{vrl}} & Virial\\
	{\texttt{pot }} & Potential energy\\
	{\texttt{r\_ctf }} & cutoff radius ($r_c$)\\
	{\texttt{U}} & Lennard-Jones potential function\\
	{\texttt{kin}} & Kinetic energy\\
	{\texttt{v2}} & $v^2$\\
	{\texttt{k\_AVG}}& Average kinetic energy\\
	{\texttt{p\_AVG}} & Average potential energy\\
	{\texttt{etot\_AVG}} & Average total energy\\
	{\texttt{T\_i}} & Instantaneous temperature\\
	{\texttt{B}} & $\beta=\sqrt{T_{0}/T}$\\
\end{tabular}

\begin{tabular}{lp{0.8\textwidth}}
	{\texttt{P}} & Pressure\\
	{\texttt{Z}} & Compressibility factor, $PV\Big/Nk_{B}T$\\
	{\texttt{step}} & Timestep counter\\	
\end{tabular}\\\\\\
To run the codes, say \texttt{code1.py}, simply write in terminal (i.e., Linux command-line interface, CLI):{\hspace{5mm}}{\texttt{python3 code1.py}}.
\subsection{MD codes in Python}
\subsubsection{code1.py}
\begin{lstlisting}[tabsize=2]
#!/usr/bin/pyhon3
import numpy as np
from math import *
import random
__d__ = 3
NS = 1000
N = 100
dt = 1.E-4
T_0 = 300.
sig = 1.
eps = 1.
r_ctf = 2.5*sig
u_at_ctf = 4.*eps*((sig/r_ctf)**12 - (sig/r_ctf)**6)
du_at_ctf = 24.*eps*(-2.*sig**12/r_ctf**13 \
               + sig**6/r_ctf**7)
bs = 10.*sig
vol = bs**3
rho = N/vol
ign = 20
#-------------------------------------------------
# Initiate positions, velocities, & accelerations
pos = np.zeros([__d__, N])
vel = np.zeros([__d__, N])
acc = np.zeros([__d__, N])
for j in range (N):
    for i in range (__d__):
        pos[i,j] = random.uniform(0, bs)
        vel[i,j] = random.uniform(0, 1)
#--------------------------------------
# Scale positions to the box size.
pos = pos/bs
#----------------------------
# Translate positions to the 
# center-of-mass (com) reference frame.
com = np.zeros ([__d__])
for i in range(__d__):
    for j in range(N):
        com[i] += pos[i, j]
        com[i] = com[i]/N
for i in range(__d__):
   pos[i, :] -= com[i]
#---------------------
P_S = 0.
k_S = 0.
p_S = 0.
T_S = 0.
f_tpz = open('tpz.out', 'w')
f_kpe = open('kpe.out', 'w')
f_xyz = open('pos.xyz', 'w')
R = np.zeros([__d__, N, N])
r = np.zeros([__d__, N, N])
#--------------------------
def sign(a, b):
    """
       The sign function
    """
    if b > 0:
       return a
    elif b < 0:
       return -a
    elif b == 0:
       return a
#------------------------------------
# The time evolution loop (main loop)
for step in range(1, NS + 1):
   #-----------------------------
   # Periodic boundary conditions
   pos[np.where(pos > 0.5)] -= 1
   pos[np.where(pos < -0.5)] += 1
   #-----------------------------
   # Loop for computing forces
   acc = np.zeros([__d__, N])
   pot = np.zeros([N])
   vrl = 0.
   for i in range(N):
        for j in range(i + 1, N):
           if j != i:
                r2 = np.zeros([N, N])
                for k in range(__d__):
                    R[k, i, j] = pos[k, i] - pos[k, j]
                    if abs(R[k, i, j]) > 0.5:
                       R[k, i, j] -= sign(1., R[k, i, j])
                    r[k, i, j] = bs*R[k, i, j]
                    r2[i, j] += r[k, i, j]*r[k, i, j]
                if r2[i, j] < r_ctf*r_ctf:
                       r1 = sqrt(r2[i, j])
                       ri2 = 1./r2[i, j]
                       ri6 = ri2*ri2*ri2
                       ri12 = ri6*ri6
                       sig6 = sig**6
                       sig12 = sig**12
                       u = 4.*eps*(sig12*ri12 - sig6*ri6) \
                       - u_at_ctf - r1*du_at_ctf
                       du = 24.*eps*ri2*(2.*sig12*ri12 - sig6*ri6) \
                       + du_at_ctf*sqrt(ri2)
                       pot[j] += u
                       vrl -= du*r2[i, j]
                       for k in range(__d__):
                           acc[k, i] += du*R[k, i, j]
                           acc[k, j] -= du*R[k, i, j]
   vrl = -vrl/__d__
   #------------------
   # Update positions
   pos += dt*vel + 0.5*acc*dt*dt
   #----------------------------
   # Compute temperature
   kin = np.zeros([N])
   v2 = np.zeros([N])
   for j in range(N):
       for i in range(__d__):
           v2[j] += vel[i, j]*vel[i, j]*bs*bs
       kin[j] = 0.5*v2[j]
   k_AVG = sum(kin)/N
   T_i = 2.*k_AVG/__d__
   B = sqrt(T_0/T_i)
   #--------------------------------
   # Rescale & update the velocities
   # according to velocity Verlet algorithm
   vel = B*vel + 0.5*dt*acc
   vel += 0.5*dt*acc
   #---------------------
   # Compute temperature
   kin = np.zeros([N])
   v2 = np.zeros([N])
   for j in range(N):
       for i in range(__d__):
           v2[j] += vel[i, j]*vel[i, j]*bs*bs
       kin[j] = 0.5*v2[j]
   k_AVG = sum(kin)/N
   T_i = 2.*k_AVG/__d__
   B = sqrt(T_0/T_i)
   #------------------
   p_AVG = sum(pot)/N
   etot_AVG = k_AVG + p_AVG
   P = rho*T_i + vrl/vol
   Z = P*vol/(N*T_i)
   f_tpz.write("%e  %e  %e  %e\n" \
   %(step*dt, T_i, P, Z))
   f_kpe.write("%e  %e  %e  %e\n" \
   %(step*dt, k_AVG, p_AVG, etot_AVG))
   #----------------------------------
   # Write position components
   f_xyz.write("%d\n\n" %(step))
   for i in range(N):
       f_xyz.write("%e  %e  %e\n" %(pos[0, i]*bs, \
       pos[1, i]*bs, pos[2, i]*bs))
   #--------------------------------
   if step > ign:
      P_S += P
      k_S += k_AVG
      p_S += p_AVG
      T_S += T_i
#------------------------
   if step%(NS*0.1) == 0: # To inform that the code is being run
      print(">>>>>  Iteration #  %d  out of %d" %(step, NS))
print('\n >>>>>  Statistical Averages  <<<<<\n')
print('     Temperature = %f K' %(T_S/(NS - ign)))
print('  Kinetic energy = %f eps' %(k_S/(NS - ign)))
print('Potential energy = %f eps' %(p_S/(NS - ign)))
print('    Total energy = %f eps' %((k_S + p_S)/(NS - ign)))
print('        Pressure = %f eps/(sig^3)\n' %(P_S/(NS - ign)))
\end{lstlisting}
\subsubsection{code2.py}
\begin{lstlisting}[tabsize=2]
#!/usr/bin/pyhon3
import random, numpy as np
from math import *
#
def PAR():
    """
       Simulation parameters
    """
    __d__ = 3
    NS = 1000
    N = 100
    dt = 1.0E-4
    T_0 = 300.
    sig = 1.
    eps = 1.
    r_ctf = 2.5*sig
    u_at_ctf = 4.*eps*((sig/r_ctf)**12 - (sig/r_ctf)**6)
    du_at_ctf = 24.*eps*(-2.*sig**12/r_ctf**13 \
                   + sig**6/r_ctf**7)
    bs = 10.*sig
    vol = bs**3
    rho = N/vol
    ign = 20
    return __d__, NS, N, dt, T_0, sig, \
    eps, r_ctf, u_at_ctf, du_at_ctf, bs, vol, rho, ign
#------------------------------------------------------
def INIT(__d__, N, bs):
    """
       Initialize positions,
       velocities, 
       and accelerations
    """
    pos = np.zeros([__d__, N])
    vel = np.zeros([__d__, N])
    acc = np.zeros([__d__, N])
    for j in range (N):
        for i in range (__d__):
            pos[i, j] = random.uniform(0, bs)
            vel[i, j] = random.normalvariate(0, 1)
    return pos, vel, acc
#------------------------
def SCL(bs, pos):
    """
       Scale positions, 
       velocities, 
       and accelerations
       to the box size
    """
    pos = pos/bs
    return pos
#-------------------------
def COMREF(__d__, N, pos):
    """
       Translate positions
       to the center-of-mass 
       (com) reference frame
    """
    com = np.zeros ([__d__])
    for i in range(__d__):
        for j in range(N):
            com[i] += pos[i, j]
            com[i] = com[i]/N
    for i in range(__d__):
        pos[i, :] -= com[i]
    return pos
#--------------
def sign(a, b):
    """
       The sign function:
       sign(a, b) returns 'a' 
       with the sign of 'b'
    """
    if b > 0:
       return a
    elif b < 0:
       return -a
    elif b == 0:
       return a
#--------------------------------------------
def FRC(pos, N, __d__, bs, sig, eps, r_ctf, \
               u_at_ctf, du_at_ctf):
    """
       Compute forces, 
       potential energy,
       and virial
    """
    acc = np.zeros([__d__, N])
    pot = np.zeros([N])
    vrl = 0.
    for i in range(N - 1):
        for j in range(i + 1, N):
            R = [pos[0, i] - pos[0, j], pos[1, i] - pos[1, j], \
            pos[2, i] - pos[2, j]]
            if abs(R[0]) > 0.5:
                R[0] -= sign(1., R[0])
            if abs(R[1]) > 0.5:
                R[1] -= sign(1., R[1])
            if abs(R[2]) > 0.5:
                R[2] -= sign(1., R[2])
            r2 = bs*bs*np.dot(R, R)
            if r2 < r_ctf*r_ctf:
                   r1 = sqrt(r2)
                   ri2 = 1./r2
                   ri6 = ri2*ri2*ri2
                   ri12 = ri6*ri6
                   sig6 = sig*sig*sig*sig*sig*sig
                   sig12 = sig6*sig6
                   u = 4.*eps*(sig12*ri12 - sig6*ri6) \
                   - u_at_ctf - r1*du_at_ctf
                   du = 24.*eps*ri2*(2.*sig12*ri12 - sig6*ri6) \
                   + du_at_ctf/r1
                   pot[j] += u
                   vrl -= du*r2
                   for k in range(__d__):
                       acc[k, i] += du*R[k]
                       acc[k, j] -= du*R[k]
    return acc, pot, vrl
#--------------------------------
def TEMP(vel, N, __d__, bs, T_0):
    """
       Compute instantaneous temperature
    """
    kin = np.zeros([N])
    v2 = np.zeros([N])
    for j in range(N):
        for i in range(__d__):
            v2[j] += vel[i, j]*vel[i, j]*bs*bs
        kin[j] = 0.5*v2[j]
    k_AVG = np.mean(kin)
    T_i = 2.*k_AVG/__d__
    B = sqrt(T_0/T_i)
    return T_i, B, k_AVG
#----------------------------------------------
def EVOL(N, __d__, NS, pos, vel, acc, dt, bs, \
                   sig, eps, r_ctf, u_at_ctf, \
                   du_at_ctf, T_0, rho, vol, ign):
    P_S = 0.
    k_S = 0.
    p_S = 0.
    T_S = 0.
    f_tpz = open('tpz.out', 'w')
    f_kpe = open('kpe.out', 'w')
    f_xyz = open('pos.xyz', 'w')
    for step in range(1, NS + 1):
       #----------------------------
       pos[np.where(pos > 0.5)] -= 1
       pos[np.where(pos < -0.5)] += 1
       #-----------------------------------
       acc, pot, vrl = FRC(pos, N, __d__, \
       bs, sig, eps, r_ctf, u_at_ctf, du_at_ctf)
       vrl = -vrl/__d__
       #---------------
       """
          Updating positions
       """
       pos += dt*vel + 0.5*acc*dt*dt
       T_i, B, k_AVG = TEMP(vel, N, __d__, bs, T_0)
       """
          Update velocities
       """
       vel = B*vel + 0.5*dt*acc
       acc, pot, vrl = FRC(pos, N, __d__, \
       bs, sig, eps, r_ctf, u_at_ctf, du_at_ctf)
       vel += 0.5*dt*acc
       T_i, B, k_AVG = TEMP(vel, N, __d__, bs, T_0)
       p_AVG = np.mean(pot)
       etot_AVG = k_AVG + p_AVG
       P = rho*T_i + vrl/vol
       Z = P*vol/(N*T_i)
       f_tpz.write("%d  %e  %e  %e  %e\n" \
       %(step, step*dt, T_i, P, Z))
       f_kpe.write("%d  %e  %e  %e  %e\n" \
       %(step, step*dt, k_AVG, p_AVG, etot_AVG))
       f_xyz.write("%d\n\n" %(step))
       for i in range(N):
           f_xyz.write("%e  %e  %e\n" \
           %(pos[0, i]*bs, pos[1, i]*bs, pos[2, i]*bs))
       if step > ign:
          """
             A condition for ignoring 
             the first "ign" steps 
             in order for reliable 
             statistical averaging.
          """
          P_S += P
          k_S += k_AVG
          p_S += p_AVG
          T_S += T_i
       if step%(NS*0.1) == 0: # To inform that the code is being run
           print(">>>>>  Iteration #  %d  out of %d" %(step, NS))
    return P_S, k_S, p_S, T_S, \
           f_tpz, f_kpe, f_xyz
#------------------------------
def STAVG(NS, ign, P_S, k_S, \
    p_S, T_S, f_tpz, f_kpe, f_xyz):
    """
       Write statistical averages
    """
    f_AVG = open('means.out', 'w')
    f_AVG.write('>>>>>  Statistical Averages  <<<<<\n\n')
    f_AVG.write('     Temperature = %f K\n' %(T_S/(NS - ign)))
    f_AVG.write('  Kinetic energy = %f eps\n' %(k_S/(NS - ign)))
    f_AVG.write('Potential energy = %f eps\n' %(p_S/(NS - ign)))
    f_AVG.write('    Total energy = %f eps\n' %((k_S + p_S)/(NS - ign)))
    f_AVG.write('        Pressure = %f eps/(sig^3)\n' %(P_S/(NS - ign)))
    f_AVG.close()
    f_tpz.close()
    f_kpe.close()
    f_xyz.close()
    print('\n>>>>>  Statistical Averages  <<<<<\n')
    print('     Temperature = %f K' %(T_S/(NS - ign)))
    print('  Kinetic energy = %f eps' %(k_S/(NS - ign)))
    print('Potential energy = %f eps' %(p_S/(NS - ign)))
    print('    Total energy = %f eps' %((k_S + p_S)/(NS - ign)))
    print('        Pressure = %f eps/(sig^3)\n' %(P_S/(NS - ign)))
    return f_AVG
#----------------------------
__d__, NS, N, dt, T_0, sig, \
eps, r_ctf, u_at_ctf, du_at_ctf, bs, \
vol, rho, ign = PAR()
#---------------------------------
pos, vel, acc = INIT(__d__, N, bs)
pos = SCL(bs, pos)
pos = COMREF(__d__, N, pos)
#--------------------------
P_S, k_S, p_S, T_S, \
f_tpz, f_kpe, f_xyz = EVOL(N, \
__d__, NS, pos, vel, acc, dt, bs, sig, eps, \
r_ctf, u_at_ctf, du_at_ctf, T_0, \
rho, vol, ign)
#----------------------
f_AVG = STAVG(NS, ign, \
P_S, k_S, p_S, T_S, \
f_tpz, f_kpe, f_xyz)
\end{lstlisting}
\section{\label{sec:5}Conclusions}
A brief, comprehensive overview of MD simulation was provided which encompasses all the required fundamentals making this introduction self-contained. The associated computer code, simulating a gaseous system composed of classical Lennard-Jones particles was also written in Python programming language.
\appendix
\section{Equivalence between Verlet and velocity Verlet algorithms}
\label{sec:app1}
From Eq.~\ref{eq:a1}, we have
\begin{eqnarray*}
	x_{n+1}=2x_{n}-x_{n-1}+a_{n}h^{2}=x_{n}+\frac{1}{2}(x_{n+1}-x_{n-1})-\frac{1}{2}x_{n-1}-\frac{1}{2}x_{n+1}+x_{n}+a_{n}h^{2}\nonumber\\
	=x_{n}+v_{n}h-\frac{1}{2}a_{n}h^{2}+a_{n}h^{2}=x_{n}+v_{n}h+\frac{1}{2}a_{n}h^{2}.
\end{eqnarray*}
Also, in Eq.~\ref{eq:vervelv}, the last term is indeed the area of a trapezoid in trapezoidal integration method, which is a more better approximation compared to the rectangular analogue. This term could also be viewed as the time average of the acceleration.
\section{Equivalence between single- and double-particle virial terms}
\label{sec:app2}
\begin{figure}[H]
	\centering
	\fbox{\rule[0cm]{0cm}{0cm} \rule[0cm]{0cm}{0cm}
	\includegraphics[scale=0.3]{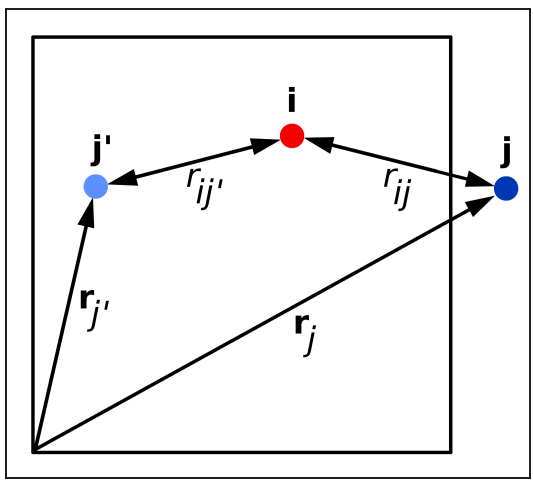}}
	\caption{\label{fig:8}
		The simulation box (primary cell) containing the two particles i and j. As the latter leaves the box, its image, j', simultaneously enters from the opposite side. The important point is that while single-particle positions ${\mathrm{\bf{r}}}_{j}$ and ${\mathrm{\bf{r}}}_{j'}$ are very different compared to each other, the double-particle ones, namely $r_{ij}$ and $r_{ij'}$, are nearly the same.}
\end{figure}

We have
\begin{eqnarray*}
	\sum_{i}\mathrm{\bf{F}}_{i}.\mathrm{\bf{r}}_{i}=\sum_{i}\Bigg(\sum_{j}\mathrm{\bf{F}}_{ij}\Bigg).\mathrm{\bf{r}}_{i}=\sum_{i}\Bigg(\frac{1}{2}\sum_{j}(\mathrm{\bf{F}}_{ij}-\mathrm{\bf{F}}_{ji})\Bigg).\mathrm{\bf{r}}_{i}
	=\frac{1}{2}\sum_{i}\sum_{j}\mathrm{\bf{F}}_{ij}.\mathrm{\bf{r}}_{i}-\frac{1}{2}\sum_{i}\sum_{j}\mathrm{\bf{F}}_{ji}.\mathrm{\bf{r}}_{i}\\
	=\frac{1}{2}\sum_{i}\sum_{j}\mathrm{\bf{F}}_{ij}.\mathrm{\bf{r}}_{i}-\frac{1}{2}\sum_{j}\sum_{i}\mathrm{\bf{F}}_{ij}.\mathrm{\bf{r}}_{j}
	=\frac{1}{2}\sum_{i}\sum_{j}\mathrm{\bf{F}}_{ij}.\mathrm{\bf{r}}_{i}-\frac{1}{2}\sum_{i}\sum_{j}\mathrm{\bf{F}}_{ij}.\mathrm{\bf{r}}_{j}
	=\frac{1}{2}\sum_{i}\sum_{j}\mathrm{\bf{F}}_{ij}.(\mathrm{\bf{r}}_{i}-\mathrm{\bf{r}}_{j})\\=\sum_{i<j}\mathrm{\bf{F}}_{ij}.\mathrm{\bf{r}}_{ij}.
\end{eqnarray*}
Nevertheless, PBC has a considerably less impact on $\mathrm{\bf{r}}_{ij}$ compared to the single-particle position $\mathrm{\bf{r}}_{i}$, as illustrated in Fig.~\ref{fig:8}.
As seen in Fig.~\ref{fig:8}, there is a marked difference between single-particle positions ${\mathrm{\bf{r}}}_{j}$ and ${\mathrm{\bf{r}}}_{j'}$, in contrast to the double-particle ones $r_{ij}$ and $r_{ij'}$, which are nearly the same.

\end{document}